# Viscosity of Silica and Doped Silica Melts: Evidence for a Crossover Temperature

John C. Mauro[1,*], Charles R. Kurkjian[2], Prabhat K. Gupta[3], and Walter Kob[4]

[1]*Department of Materials Science and Engineering, The Pennsylvania State University, University Park, Pennsylvania 16802, USA*
[2]*Department of Materials Science and Engineering, Rutgers University, Piscataway, New Jersey 08855, USA*
[3]*Department of Materials Science and Engineering, The Ohio State University, Columbus, Ohio 43210, USA*
[4]*Laboratoire Charles Coulomb, University of Montpellier and CNRS, F-34095 Montpellier, France*
[*]Corresponding author: jcm426@psu.edu

**ABSTRACT**

Silica is known as the archetypal strong liquid, exhibiting an Arrhenius viscosity curve with a high glass transition temperature and constant activation energy. However, given the ideally isostatic nature of the silica network, the presence of even a small concentration of defects can lead to a significant decrease in both the glass transition temperature and activation energy for viscous flow. To understand the impact of trace level dopants on the viscosity of silica, we measure the viscosity-temperature curves for seven silica glass samples having different impurities, including four natural and three synthetic samples. Depending on the type of dopant, the glass transition temperature can vary by nearly 300 K. A common crossover is found for all viscosity curves around ~2200-2500 K, which we attribute to a change of the transport mechanism in the melt from being dominated by intrinsic defects at high temperature to dopant-induced defects at low temperatures.

## I. Introduction

The viscosity of glass-forming systems is critically important for all stages of glass processing [1]-[3]. As such, fundamental understanding of the temperature and composition dependence of viscosity has been identified as a grand challenge in the field [4]-[5]. Following the pioneering work of Angell [6]-[7], the viscosity of glass-forming liquids can be categorized as either strong or fragile. Strong liquids exhibit an Arrhenius dependence of viscosity on temperature, whereas fragile liquids demonstrate a departure from Arrhenius behavior. The Arrhenius scaling of strong liquids is indicative of a single activation energy governing the viscous flow process.

While most glass-forming liquids exhibit some degree of fragility, the archetypal strong liquid is silica ($SiO_2$). Vitreous silica consists of a topologically disordered network of corner-sharing $(SiO_4)^{4-}$ tetrahedra [8]. Considering each tetrahedron to be a rigid polytope, i.e., with no internal degrees of freedom, the disordered silica network is nominally isostatic, i.e., where the number of rigid constraints is equal to the number of atomic degrees of freedom in three spatial dimensions [9]. In the low-temperature glassy state, the only atomic degrees of freedom arise from trace dopants or intrinsic defects such as mis-coordinated atoms [10]. Following temperature-dependent constraint theory [11]-[13], topological degrees of freedom can become activated at sufficiently high temperatures, e.g., when the temperature is high enough to allow for internal flexibility in the $(SiO_4)^{4-}$ tetrahedra. For silica, such internal flexibility would become activated only when there is enough thermal energy to break the siloxane (-Si-O-Si-) linkages.

Owing to this high degree of network rigidity, silica has one of the highest viscosities of any known glass-forming system [14]-[23]. Given the very low flexibility of the silica network, the presence of even small number of defects can have a significant effect on the viscosity-temperature relationship [15]-[20]. This follows directly from temperature-dependent constraint theory, since the glass transition temperature scales inversely with the number of atomic degrees of freedom in the network [11]-[13]. For a truly isostatic network, the number of degrees of freedom is zero. Hence, the presence of even small concentrations of defects can have an outsize effect on the viscosity curve.

Experimental investigations of silica viscosity are complicated by the high temperatures required to measure flow. Two approaches have been demonstrated to be especially effective at measuring the high-temperature viscosity of silica, viz., the fiber drawing [15] and torsional [16] techniques. In this paper, we employ both methods to conduct a systematic investigation of the impact of dopants on silica viscosity. We consider seven different types of silica melts, including both natural and synthetic samples, revealing the impact of dopants on the resulting glass transition temperature and activation energy for viscous flow. Interestingly, our measurements suggest the existence of a crossover at finite temperature for the viscosity of the doped samples. We provide an estimate of the crossover temperature (which is higher than all measurement temperatures) and discuss its significance.

**II. Experimental Procedure**

The silica glass samples were produced by Heraeus Quarzglas [21] through different production processes and differ with respect to impurity concentrations, as described in Table I. The first four samples (LA, LS, HERASIL, and TSL) are Type II silica, i.e., produced via melting of natural quartz in a gas flame [1]. The LA and LS samples have been purified to reduce the alkali dopant concentrations. All of the Type II silica samples are "wet," i.e., with approximately 150 ppm of OH. The remaining three samples (F300, F310, and F320) are synthetic silica glass samples produced by reacting silicon tetrachloride with oxygen in a multi-step process involving soot deposition and consolidation [1]. F310 is Type III silica glass, i.e., produced in an OH-containing flame. F300 and F320 are both Type IV (dry) silica glass, which used either chlorine (F300) or fluorine (F320) gas as the drying agent [21]. Hence, the compositional variables under study are the presence or absence of alkali, OH, and halide impurities.

High-temperature viscosity measurements were conducted at AT&T Bell Laboratories using both the fiber drawing [15] and torsion [16] techniques to cover a wide temperature range of 1273-2573 K (1000-2300°C). The fiber draw technique was used to obtain the viscosity data in the upper temperature regime (1973-2573 K). The torsion method was used to measure the viscosity between 1273 and 2273 K. The two methods employ different measurement principles based on elongation and torsion, as described in detail by Paek et al. [15] and Weiss [16], respectively.

Test rods were prepared for the viscosity measurements by collapsing as-received silica tubes (25 mm × 19 mm) for waveguide applications using oxyhydrogen burners, with additional surface treatment to reduce the OH content at the outer surface. The collapsed rod diameter was ~14 mm. With the fiber drawing method, the 14-mm diameter collapsed rods were drawn into 125-µm fiber at a speed of 1 m/s. The torsional method requires a smaller sample diameter and, hence, the collapsed rods were drawn down to approximately 2-mm diameter inside a graphite furnace. The furnace temperature was set to be 2123 K (1850°C), and prior to drawing an additional surface treatment was again applied to reduce the OH content at the outer surface.

## III. Results and Discussion

The results from both the fiber and torsion techniques are described well using an Arrhenius relation,

$$\eta(T) = \eta_\infty \exp\left(\frac{E}{RT}\right), \tag{1}$$

where $\eta$ is the shear viscosity, $E$ is the activation energy for viscous flow, $R$ is the gas constant, $T$ is absolute temperature, and $\eta_\infty$ is the extrapolated infinite-temperature limit of viscosity. The data for all samples are linear on a log $\eta$ vs. $1/T$ plot, with a deviation from linearity less than 2.5% for all samples. Figure 1 shows an example Arrhenius plot for the LA sample. The error in the individual viscosity measurements is less than 0.1 logarithmic units, i.e., smaller than the size of the symbols in the plot. The data in Figure 1 are overlaid with the previously published viscosity data for various Type II silicate glasses, as compiled by Nascimento and Zanotto [18].

The Arrhenius parameters for all samples are listed in Table II, where the discrepancy between the two measurement techniques is evaluated in terms of the difference in calculated values of $E$ and the glass transition temperature ($T_g$), which are fitted using Eq. (1). Following Angell [6]-[7], the glass transition temperature, $T_g$, is defined as the temperature at which the shear viscosity is equal to $10^{12}$ Pa·s. As can be seen in Table II, the agreement between the two techniques is excellent, except for the F320 glass where the discrepancy is 21% in $E$ and 6.9% in $T_g$. This is an extreme case, as the normal discrepancy is less than ~8% or ~4% for $E$ and $T_g$, respectively. The Arrhenius viscosity curves are shown in Figure 2.

Let us first consider the results obtained for the natural Type II glasses, viz., LA, LS, Herasil, and TSL. Although the variations in $E$ are somewhat greater than the estimated experimental error, the values found with the individual measurement techniques are essentially the same, ~580 kJ/mol for fiber drawing and ~600 kJ/mol for the torsional technique, since OH is the dominant dopant in each of these glasses, with a similar OH concentration in all four samples. The higher values of both $E$ and $T_g$ for the higher purity LA and LS glasses are understandable in terms of intrinsic defects, which according to Mott [24], represent neighboring pairs of broken siloxane linkages. The low values of $E$ and $T_g$ for the high Cl and F glasses (F300 and F320) are also expected due to the presence of extrinsic defects representing chemically induced broken siloxane linkages. The $E$ and $T_g$ values for the Type III F310 glass are slightly higher than those of the F300 and F320 glasses due to its lower halide concentration.

As shown in Figure 2, there is an apparent crossover of the viscosity curves in the range of 2200-2500 K. The existence of a finite crossover temperature, $T_c$, leads to a linear relationship between $1/T_g$ and $1/E$,

$$\frac{R\ln 10\left[12-\log\eta(T_c)\right]}{E} = \frac{1}{T_g} - \frac{1}{T_c}. \qquad (2)$$

This linearity is confirmed in Figure 3, which provides further evidence for the existence of this crossover temperature. Some evidence for a crossover temperature can also be seen in the silica viscosity data compiled by Nascimento and Zanotto [18] (see their Fig. 6).

The value of the crossover temperature, $T_c$, can be determined from the slope of the best-fit lines in Figure 4 via:

$$T_c = -\frac{1}{\dfrac{d\log\eta_\infty}{dE}R\ln 10}. \qquad (3)$$

From Figure 4 and Eq. (3), $T_c$ is found to be ~2511 K for the fiber draw method and ~2223 K for the torsional technique. This range of temperatures corresponds to viscosities of about $10^{3.5}$-$10^{5.5}$ Pa·s. This range falls between the softening point ($10^{6.6}$ Pa·s) and the working point ($10^3$ Pa·s) of the glass-forming liquid, indicating that all of the samples should exhibit a similar working point.

The high activation energy (~600 kJ/mol) of the silica viscosity curve is considered to be a consequence of formation of thermally activated intrinsic defects which according to Mott represent neighboring pairs of broken Si-O-Si network linkages [24]. Small amounts (2000 ppm or less) of network modifying dopants (such as OH, Cl, F, and Na) also cause breaks in network siloxane (-Si-O-Si-) linkages, leading to formation of non-thermally activated extrinsic defects in the network. The effect of these dopants is a reduction in the activation energy, while the behavior of doped silica liquids remains Arrhenius (i.e., strong). Computer simulations by Horbach and Kob [25] and by Geske et al. [26] show that at high temperatures (much higher than accessible in viscosity experiments), silica becomes fragile, exhibiting a so-called fragile-to-strong transition in the range of 3100–3300 K. Experimental verification of high temperature fragile behavior is not yet possible due to limitations in high temperature viscosity measurements. Higher temperature measurement would be required to probe this question directly, for example, through aerodynamic levitation [27].

The origin of the apparent crossover of the viscosity curves around 2200-2500 K is explained by the physically sound assertion that the viscosity curves of doped silicas cannot be higher than that of pure silica and must transition to that of pure silica at higher temperatures, i.e., where intrinsic defects become dominant. Hence, we expect that for temperatures above $T_c$, the viscosities of the different systems considered become to a first approximation independent of the amount of doping.

**IV. Conclusions**

We have investigated the viscosity-temperature relationship for seven silica glass samples at high temperatures. Among the seven samples, four were produced from natural quartz and three were synthetically fabricated through vapor deposition. Reasonably good agreement is found for the fiber drawing and torsional techniques for measuring viscosity. The viscosity curves of the four natural Type II glasses (LA, LS, Herasil, and TSL) are similar. On the other hand, the three synthetic glasses (F300, F310, and F320), which contain varying amounts of OH, Cl, and F, are quite different. The Type III silica glass having ~150 ppm OH contamination (F310) exhibited higher glass transition temperature and activation energy for viscous flow compared to the two dry Type IV samples (F300 and F320), which had ~150 ppm of halide contamination. The fluorine-doped Type IV sample has lower $T_g$ and $E$ compared to the chlorine-doped sample. Interestingly, all seven samples appear to exhibit, upon extrapolation, common viscosity values around 2200-2500 K, which may be attributed to intrinsic defects becoming dominant at high temperatures. The value of $T_c$, in principle, may change with the dopant level. However, for the compositions studied, this change in $T_c$ is not significant, as shown by the straight line fit in Figure 4.

**Acknowledgments**

C.R. Kurkjian is grateful to his former colleagues from Bell Labs, U.C. Paek, C.M. Schroeder, K. Klein, W. Weiss, and C. Schmitt, and all their technical support and collaboration. The authors would like to thank M.L.F. Nascimento and E.D. Zanotto for sharing their compilation of viscosity data.

**Figure Captions**

**Figure 1.** Viscosity curve for the Type II LA SiO$_2$ sample measured by the fiber drawing (open circles) and torsional (closed circles) methods. Error bars are smaller than the size of the symbols. The gray crosses show the literature data for Type II silica, as compiled by Nascimento and Zanotto [18].

**Figure 2.** Arrhenius viscosity-temperature relationships for the seven types of silica, interpolated and extrapolated from (a) fiber drawing and (b) torsional measurements. For the torsional technique, the crossover temperature shifts to a somewhat lower value compared to the fiber draw technique.

**Figure 3.** $1/T_g$ vs. $1/E$ for the (a) fiber drawing and (b) torsional measurements. The linearity is the consequence of the existence of a crossover point at a finite temperature, $T_c$.

**Figure 4.** Plot of -log $\eta_\infty$ vs. $E$ for the silica glasses measured through both the fiber drawing and torsional methods. The best-fit lines are used to obtain $T_c$ values of 2223 K from the torsional method and 2511 K from the fiber drawing method.

**Table I.** Silica samples used in this study.

| Sample | Type | Description | Alkali (ppm) | OH (ppm) | Cl (ppm) | F (ppm) |
|---|---|---|---|---|---|---|
| LA | II | Natural Quartz, Torch, Fused, "Purified" | Low | ~150 | - | - |
| LS | II | Natural Quartz, Torch, Fused, "Purified" | Low | ~150 | - | - |
| HERASIL | II | Natural Quartz, Torch, Fused | ~100 | ~150 | - | - |
| TSL | II | Natural Quartz, Torch, Fused | ~100 | ~150 | - | - |
| F300 | IV | Hydrolyze $SiCl_4$, Dry/Sinter in $Cl_2$ | - | <1 | ~1500 | - |
| F310 | III | No Dry | - | ~150 | - | - |
| F320 | IV | Sinter in $F_2$ | - | <1 | - | ~1500 |

**Table II.** Viscosity data by fiber draw and torsional techniques. Errors are estimated to be within ±5 kJ/mol and ±5 K, respectively, for $E$ and $T_g$.

| Sample | Fiber Draw Technique | | | Torsional Technique | | | % Difference | |
|---|---|---|---|---|---|---|---|---|
| | $E$ (kJ/mol) | $\log[\eta_\infty$ (Pa·s)] | $T_g$ (K) | $E$ (kJ/mol) | $\log[\eta_\infty$ (Pa·s)] | $T_g$ (K) | % $\Delta E$ | % $\Delta T_g$ |
| LA | 587.4 | -8.60 | 1473 | 584.5 | -8.48 | 1473 | -0.5 | 0.0 |
| LS | 588.3 | -8.54 | 1479 | 589.9 | -8.63 | 1477 | 0.3 | -0.1 |
| HERASIL | 559.0 | -8.30 | 1465 | 614.2 | -9.15 | 1500 | 9.0 | 2.4 |
| TSL | 559.8 | -8.09 | 1438 | 604.2 | -8.95 | 1490 | 8.0 | 3.6 |
| F300 | 503.3 | -6.74 | 1387 | 483.7 | -6.08 | 1380 | -4.0 | -0.5 |
| F310 | 545.6 | -7.67 | 1432 | 501.7 | -6.56 | 1394 | -8.0 | -2.7 |
| F320 | 495.8 | -6.80 | 1361 | 392.5 | -3.96 | 1267 | -21.0 | -6.9 |

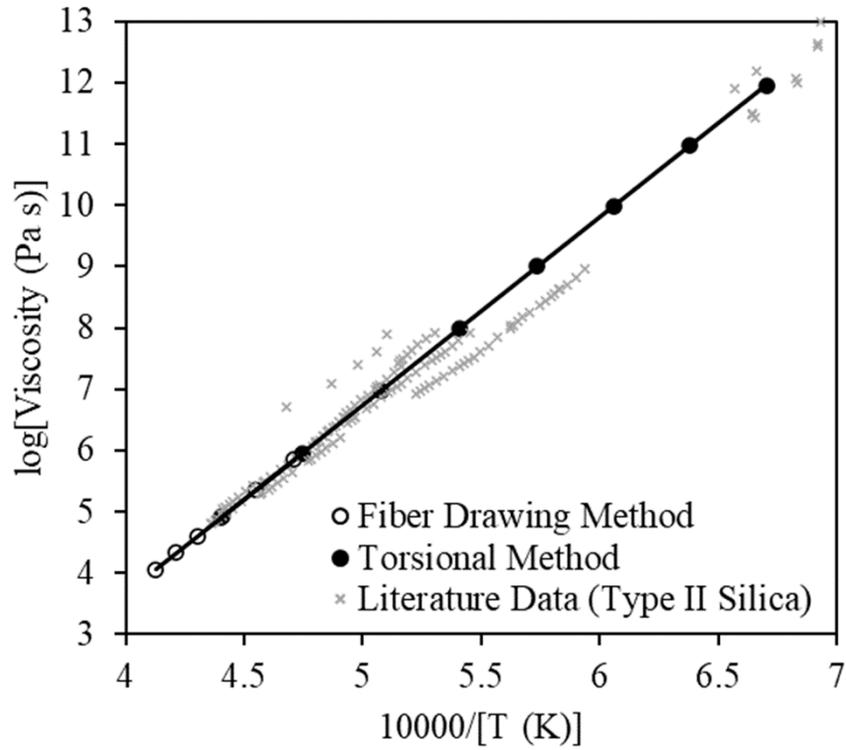

**Figure 1.** Viscosity curve for the Type II LA SiO$_2$ sample measured by the fiber drawing (open circles) and torsional (closed circles) methods. Error bars are smaller than the size of the symbols. The gray crosses show the literature data for Type II silica, as compiled by Nascimento and Zanotto [18].

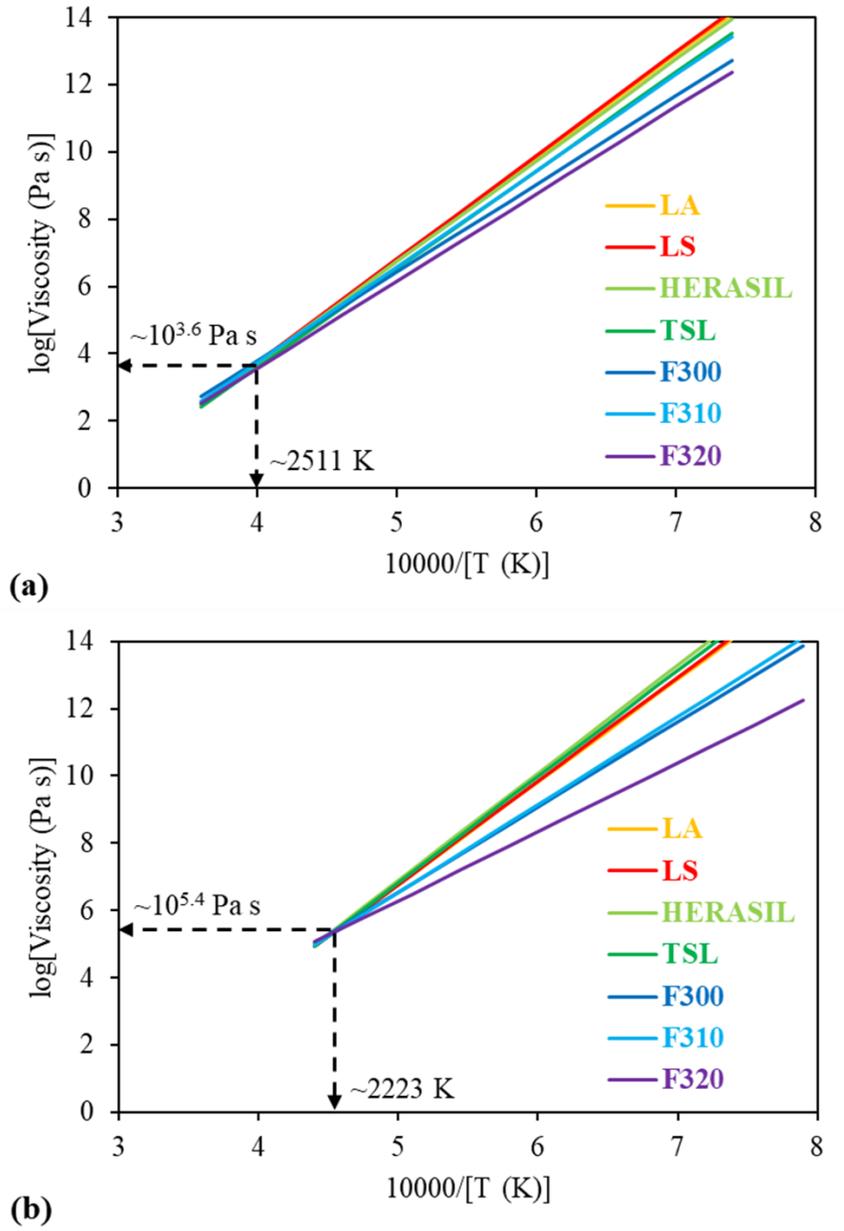

**Figure 2.** Arrhenius viscosity-temperature relationships for the seven types of silica, interpolated and extrapolated from (a) fiber drawing and (b) torsional measurements. For the torsional technique, the crossover temperature shifts to a somewhat lower value compared to the fiber draw technique.

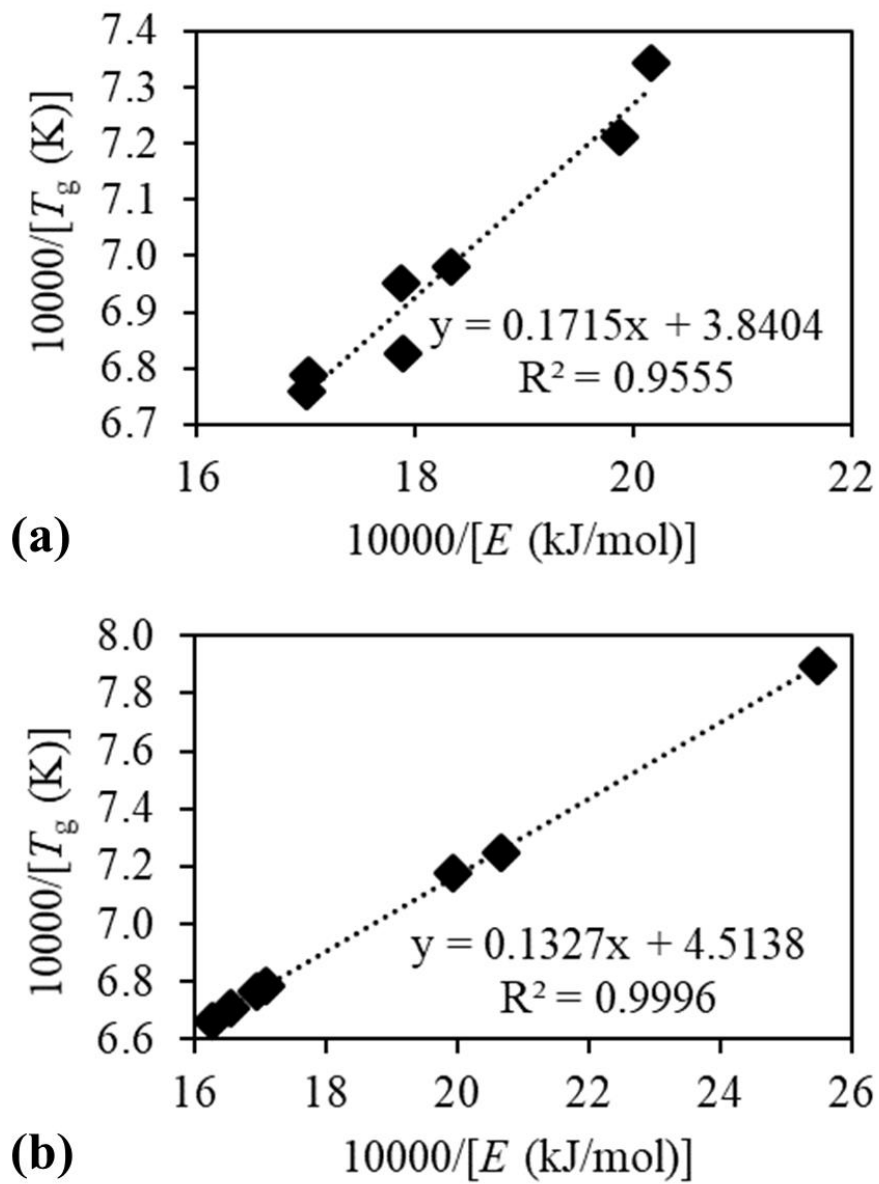

**Figure 3.** $1/T_g$ vs. $1/E$ for the (a) fiber drawing and (b) torsional measurements. The linearity is the consequence of the existence of a crossover point at a finite temperature, $T_c$.

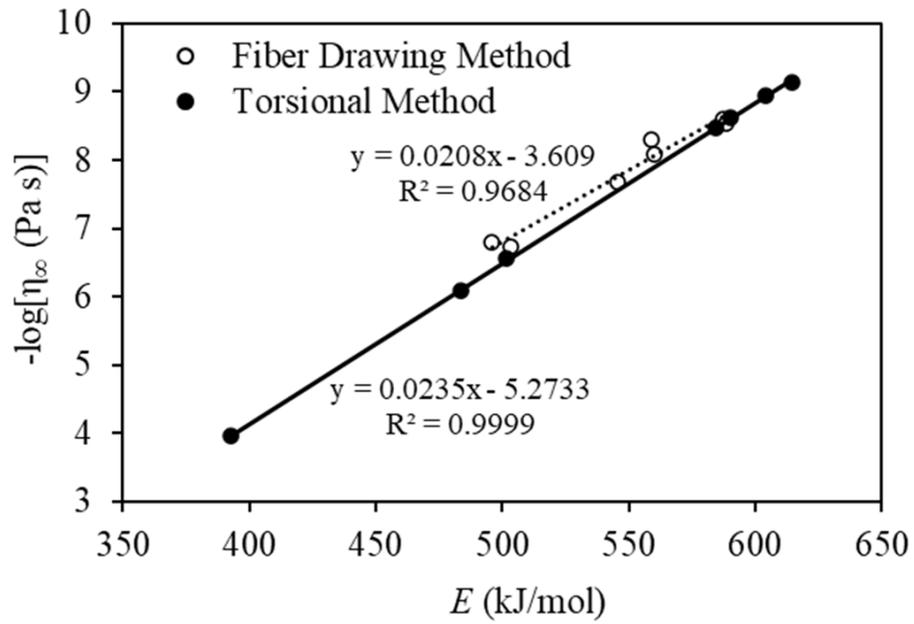

**Figure 4.** Plot of -log $\eta_\infty$ vs. $E$ for the silica glasses measured through both the fiber drawing and torsional methods. The best-fit lines are used to obtain $T_c$ values of 2223 K from the torsional method and 2511 K from the fiber drawing method.